\documentclass{article}
\usepackage[utf8]{inputenc}
\usepackage[T1]{fontenc}
\usepackage{url}
\usepackage{nicefrac}
\usepackage{microtype}

\usepackage{subcaption}
\usepackage{amsmath,graphicx,multicol}
\usepackage{amsfonts}
\usepackage{color}
\usepackage{bbm}
\usepackage{natbib}
\usepackage{amssymb}
\usepackage{algorithm}
\usepackage{algorithmic}
\usepackage{tabularx}
\usepackage[english]{babel}

\usepackage{amsthm}
\usepackage{mathtools}
\usepackage{comment}
\usepackage{array,booktabs,dcolumn,calc}
\usepackage{float}
\usepackage{colortbl}

\definecolor{rcolor}{rgb}{1,0.85,0.85}
\usepackage[keeplastbox]{flushend}

\newtheorem{definition}{Definition}[]
\newtheorem{proposition}{Proposition}[]
\newtheorem{theorem}{Theorem}[]

\def\v{{\boldsymbol{\nu}}}
\def\tt{{\boldsymbol{\theta}}}
\def\uu{{\boldsymbol{\mu}}}

\def\x{{\mathbf x}}

\def\a{{\mathbf a}}

\def\y{{\mathbf y}}

\def\z{{\mathbf z}}

\def\A{{\mathbf A}}
\def\B{{\mathbf B}}
\def\P{{\mathbf P}}
\def\M{{\mathbf M}}
\def\X{{\mathbf X}}
\def\I{{\mathbf I}}

\def\I{{\mathbf I}}
\def\F{{\mathbf F}}

\def\G{{\mathbf \Gamma}}
\def\Tt{{\mathbf \Theta}}

\def\S{{\cal S}}
\def\T{{\cal T}}
\def\Xs{{\cal X}}

\def\N{{\mathcal{N}}}

\newcommand{\ts}{\textsuperscript}

\DeclareMathOperator{\E}{\mathbb{E}}

\newcommand{\Cov}{\mathrm{Cov}}

\def\R{{\mathbb{R}}}
\def\Tr{{\mathrm{Tr}}}

\newcommand{\RNum}[1]{\uppercase\expandafter{\romannumeral #1\relax}}

\newcolumntype{C}{>{\centering\arraybackslash}b{\widthof{positions}}}
\newcolumntype{d}{D{.}{.}{-2}}

\usepackage{microtype}
\usepackage{graphicx}
\setcitestyle{nosort}
\usepackage[accepted]{icml2019}

\icmltitlerunning{Submodular Observation Selection and Information Gathering for Quadratic Models}

\begin{document}
\twocolumn[
\icmltitle{Submodular Observation Selection and Information Gathering\\for Quadratic Models}

\icmlsetsymbol{equal}{*}

\begin{icmlauthorlist}
	\icmlauthor{Abolfazl Hashemi}{1}
	\icmlauthor{Mahsa Ghasemi}{1}
	\icmlauthor{Haris Vikalo}{1}
	\icmlauthor{Ufuk Topcu}{1}
\end{icmlauthorlist}

\icmlaffiliation{1}{University of Texas at Austin}
\icmlcorrespondingauthor{Abolfazl Hashemi}{abolfazl@utexas.edu}
\icmlcorrespondingauthor{Mahsa Ghasemi}{mahsa.ghasemi@utexas.edu}

\icmlkeywords{Observation Selection, Submodular Optimization, Greedy Algorithms}
\vskip 0.3in
]
\printAffiliationsAndNotice{}

\begin{abstract}
		We study the problem of selecting most informative subset of a large observation set to enable accurate estimation of unknown parameters. This problem arises in a variety of settings in machine learning and signal processing including feature selection, phase retrieval, and target localization. Since for quadratic measurement models the moment matrix of the optimal estimator is generally unknown, majority of prior work resorts to approximation techniques such as linearization of the observation model to optimize the alphabetical optimality criteria of an approximate moment matrix. Conversely, by exploiting a connection to the classical Van Trees' inequality, we derive new alphabetical optimality criteria without distorting the relational structure of the observation model. We further show that under certain conditions on parameters of the problem these optimality criteria are monotone and (weak) submodular set functions. These results enable us to develop an efficient greedy observation selection algorithm uniquely tailored for quadratic models, and provide theoretical bounds on its achievable utility. 
\end{abstract}
\section{Introduction}\label{sec:intro}
In many machine learning applications, one needs to 
efficiently collect the most informative observations from a potentially significantly larger set of uncertain observations. The goal of such selection is to reduce the burden on computational and communication resources while still providing accurate inference of unknown parameters. For instance, to reduce the 
computational and communication costs in sensor and feature selection 
applications \cite{joshi2009sensor, shamaiah2010greedy, bhaskara2016greedy}, the goal is to select a 
small representative subset of observations gathered by different sensing units 
within a network of autonomous systems. In sensor placement 
\cite{krause2008near}, the problem is to select a subset of locations in an environment such that the measurements collected at those locations enable effective detection of unknown targets. In Bayesian experimental design, the objective is to select a subset of experiments from the set of all possible experiments to optimize a statistical metric or expected utility defined for a specific task such as inference or prediction of some parameters \cite{chaloner1995bayesian,wang2017computationally,chamon2017approximate}. This selection is motivated by the desire to reduce experimental cost and abide to limited resources.

The example applications above belong to a family of problems referred to as observation selection or information gathering \cite{krause2007near}. They can be cast as optimization problems with an objective of maximizing the information of the selected observations subject to cardinality constraints. Since such optimization problems are generally NP-hard \cite{williamson2011design,krause2014submodular}, one often resorts to heuristic schemes that find a sub-optimal subset of observations. It has been shown that, when the measurement model is linear, typical objective functions (e.g., expected utility) are scalarization of the error covariance matrix and possess a diminishing return property known as submodularity or weak (i.e. approximate) submodularity. For such objectives, a simple greedy approximation scheme achieves near-optimal observation selections with provable performance guarantees \cite{nemhauser1978analysis,krause2014submodular}. Furthermore, since expected utilities are scalarizations of the error covariance matrix, the selection criteria for linear models have a strong connection to mean square error (MSE), the desired performance metric in many applications, including those considered in this paper. These two properties, that is (weak) submodularity of typical objective functions as well as a strong connection to MSE, makes linear models appealing. In fact when the measurement model is nonlinear, one typically resort to linearization of the model or Monte-Carlo methods to find an approximate utility prior to the actual observation selection step \cite{chaloner1995bayesian,sebastiani2000maximum,flaherty2006robust,krause2008robust,wang2016efficient,davidian2017nonlinear}. However, theoretical guarantees for the performance of greedy algorithms hold only for the linearized model, i.e., for the linear approximation of the actual nonlinear model. More importantly, due to approximation, the connection to MSE (if any) becomes less explicit and hence the selected subset of observations is generally not necessarily the most informative collection of observations.

A practically appealing class of inference tasks are those described by quadratic measurement models and inverse problems that occur in many natural phenomena and real-world applications. For instance, in object tracking and localization in robotics and autonomous systems, the range measurements gathered by radar systems follow a quadratic relation \cite{skolnik1970radar,hightower2001location}. In phase retrieval applications, where the goal is to recover an unknown object from its magnitude measurements, the relation between the unknown parameter and the magnitude measurements is described by a quadratic model \cite{fienup1982phase,shechtman2015phase}.

In this paper, we consider observation selection under models where the relation between the unknown parameters and the measurements follows a quadratic mapping with additive noise. By establishing a connection between the classical Van Trees' inequality \cite{van2004detection} and alphabetical optimality criteria \cite{chaloner1995bayesian}, we devise new objective functions that exploit the quadratic relation of the observation model. Since the proposed objectives are build upon Van Trees' inequality, they enjoy an intuitive connection to MSE. In particular, by optimizing the proposed functions, one attempts to minimize a lower bound on MSE of any estimator of the unknown parameters. We further prove that these utility functions are monotone and (weak) submodular set functions under mild conditions on the statistics of the problem and parameters of the model. These results allow us to develop a simple greedy scheme for observation selection with theoretical bounds on its achievable utility without requiring any a priori approximation step. To demonstrate efficacy of the proposed framework, we consider two applications -- multi-object tracking and phase retrieval -- and empirically verify that the subsets selected by the proposed greedy algorithm outperform approaches based on random selection, and greedy selection of observations that relies on a linearized model.
\section{Submodular Observation Selection}\label{sec:back}
Consider the linear observation model $y_i = \x_i^\top\tt+\nu_i$ where $\X \in \R^{n\times m}$ is the model parameter matrix, $\y \in \R^{n}$ is the collection of $n$ observations, $\tt \in \R^{m}$ denotes the unknown parameters, and $\nu_i \sim \N(\mathbf{0},\sigma^2_i)$ is the additive Gaussian noise. Let $\S$ denote a subset of selected observations. Assuming a normal prior  $p_\tt(\Tt)=\N(\mathbf{0},\P)$ on the unknown parameters, 
the minimum variance unbiased estimator of the parameters has a closed-form expression \cite{kay2013fundamentals}
\begin{equation}
\hat{\tt}_\S = \M_\S\left(\sum_{i\in\S}y_i\x_i\right) ,
\end{equation}
where 
\begin{equation}\label{eq:cov}
\begin{aligned}
\M_\S:=&\E[(\hat{\tt}_\S-\tt)(\hat{\tt}_\S-\tt)^\top]\\=&\left(\P^{-1}+\sum_{i\in\S}\frac{1}{\sigma^2_i}\x_i\x_i^\top\right)^{-1}
\end{aligned}
\end{equation}
 is the error covariance matrix of the estimated parameters $\hat{\tt}_\S$. Therefore, since the error covariance matrix for any subset of observations is known, one can use a suitable scalarization of the error covariance (also referred to as the moment matrix) to choose a subset of observations leading to the minimal estimation error (i.e., providing the highest information) for the inference task at hand. Typical utility functions are derived from the so-called alphabetical design criteria including A-optimality, D-optimality, E-optimality, and T-optimality \cite{chaloner1995bayesian}:
\begin{itemize}
	\item  In A-optimality, we minimize the trace of the error covariance matrix $\M_\S$, or equivalently maximize
	\begin{equation}
	f^A(\S) = \mathrm{Tr}\left(\P\right)-\mathrm{Tr}\left(\M_\S\right).
	\end{equation}
	\item  In D-optimality, we minimize the log-det of the error covariance matrix $\M_\S$, or equivalently maximize
	\begin{equation}
	f^D(\S) = \log\det\left(\M_\S^{-1}\right) - \log\det\left(\P^{-1}\right).
	\end{equation}
	\item  In E-optimality, we minimize the largest eigenvalue of the error covariance matrix $\M_\S$, $\lambda_{max}$, or equivalently maximize
	\begin{equation}
	f^E(\S) = \lambda_{min}\left(\M_\S^{-1}\right) -\lambda_{min}\left(\P^{-1}\right). 
	\end{equation}
	\item  In T-optimality, we maximize the trace of inverse of the error covariance matrix $\M_\S^{-1}$, or equivalently maximize
	\begin{equation}
	f^T(\S) = \mathrm{Tr}\left(\M_\S^{-1}\right) - \mathrm{Tr}\left(\P^{-1}\right).
	\end{equation}
\end{itemize}
If, for instance, one considers A-optimality as the utility, the task of selecting a subset of $k$ out of $n$ available observations can be formally cast as the following optimization problem
\begin{equation}\label{eq:probcard}
\begin{aligned}
&\max_{\S}\quad f^A(\S)
&\text{s.t.} \qquad |\S| \leq k.
\end{aligned}
\end{equation}

By a reduction to the set cover problem \cite{williamson2011design}, it has been shown that finding an optimal solution to \eqref{eq:probcard} is NP-hard. Nonetheless, as we mentioned in Section \ref{sec:intro}, it has been shown that observation selection and information gathering under linear models using the above optimality measures can be cast as the problem of maximizing a {\it monotone, (weak) submodular} function subject to cardinality (or uniform matroid) constraints \cite{nemhauser1978analysis}. These properties and their implications on solving the optimization problem \eqref{eq:probcard} are stated next.

\begin{definition}
\label{def:mon}
Set function $f:2^\Xs\rightarrow \mathbb{R}$  is monotone if $f(\S)\leq f(\T)$ for all $\S\subseteq \T\subseteq \Xs$.
\end{definition}

\begin{definition}
\label{def:submod}
Set function $f:2^\Xs\rightarrow \mathbb{R}$ is submodular if 
\begin{equation}
f(\S\cup \{j\})-f(\S) \geq f(\T\cup \{j\})-f(\T)
\end{equation}
for all subsets $\S\subseteq \T\subset \Xs$ and $j\in \Xs\backslash \T$. The term $f_j(\S)=f(\S\cup \{j\})-f(\S)$ is the marginal value of adding element $j$ to set $\S$.
\end{definition}

\begin{definition}
The multiplicative weak-submodularity constant of a monotone non-decreasing function $f$ is defined as
 \begin{equation}
 c_{f}={\max_{(\S,\T,i)\in \tilde{\Xs}}{f_i(\T)\slash f_i(\S)}},
 \end{equation}
where $\tilde{\Xs} = \{(\S,\T,i)|\S \subseteq \T \subset \Xs, i\in \Xs \backslash \T\}$. 
\end{definition}
The multiplicative weak-submodularity constant \cite{zhang2016submodular,chamon2017approximate} is a closely related concept to submodularity and essentially quantifies how close the set function is to being submodular. It is worth noting that a set function $f(\S)$ is submodular if and only if its multiplicative weak-submodularity constant satisfies 
$c_{f} \le 1$ \cite{das2011submodular,elenberg2016restricted,horel2016maximization}.

A similar notion of weak submodularity is the additive weak-submodularity constant of defined below \cite{zhang2016submodular,chamon2017approximate}.

\begin{definition}
The additive weak-submodularity constant of a monotone non-decreasing function $f$ is defined as
 \begin{equation}
 \epsilon_{f}={\max_{(\S,\T,i)\in \tilde{\Xs}}{f_i(\T)- f_i(\S)}},
 \end{equation}
where $\tilde{\Xs} = \{(\S,\T,i)|\S \subseteq \T \subset \Xs, i\in \Xs \backslash \T\}$. 
\end{definition}
Note that when $f(\S)$ is submodular, its additive weak-submodularity constant satisfies $\epsilon_{f}\leq 0$.

For a monotone function with bounded additive and multiplicative weak-submodularity constants (WSCs) we have the following proposition. \footnote{Detailed proofs of all lemmas, propositions, and theorems are stated in the supplementary.}
\begin{proposition}\label{lem:curv}
Let $c_f$ and $\epsilon_f$ be the multiplicative and  additive weak-submodularity constants of $f(\S)$, a monotone non-decreasing function with $f(\emptyset)=0$. Let $\S$ and $\T$ be any subsets such that $\S\subset \T \subseteq \Xs$ with $|\T\backslash \S|=r$. Then, it holds that
\begin{equation}
f(\T)-f(\S)\leq \frac{1}{r}\left(1+(r-1)c_f\right)\sum_{j\in \T\backslash \S}f_j(\S),
\end{equation}
and 
\begin{equation}
f(\T)-f(\S)\leq (r-1)\epsilon_f+\sum_{j\in \T\backslash \S}f_j(\S).
\end{equation}
\end{proposition}
\begin{algorithm}[t]
	\caption{Greedy Observation Selection}
	\label{alg:greedy}
	\begin{algorithmic}[1]
		\STATE \textbf{Input:}  Utility function $f(\S)$, set of all observations $\Xs$, number of selected observations $k$.
		\STATE \textbf{Output:} Subset $\S_g\subseteq \Xs$ with $|\S_g|=k$.
		\STATE Initialize $\S_g = \emptyset$
		\FOR{$i = 0,\dots, k-1$}
		\STATE $j_s = \text{argmax}_{j \in \Xs\backslash \S_g} f_j(\S_g)$
		\STATE $\S_g \leftarrow \S_g  \cup \{j_s\}$
		\ENDFOR
	\end{algorithmic}
\end{algorithm}
It has been shown  in \cite{nemhauser1978analysis} that if a set function is monotone and submodular, a simple greedy algorithm that iteratively select an observation with the highest marginal gain (see Algorithm 1) satisfies a $1-1\slash e$ approximation factor. Using Proposition \ref{lem:curv}, one can extend these theoretical results to the case of weak submodular functions, as illustrated in the following proposition \cite{das2011submodular,chamon2017approximate,elenberg2016restricted,hashemi2017randomized}.

\begin{proposition}\label{thm:curv}
Let $c_f$ and $\epsilon_f$ be the multiplicative and additive weak-submodularity constants of $f(\S)$, a monotone non-decreasing function with $f(\emptyset)=0$. Let $\S_g \subseteq \Xs$ with $|\S_g|\leq k$ be the subset selected when maximizing $f(\S)$ subject to a cardinality constraint  via the greedy observation selection scheme, and let $\S^\star$ denote the optimal subset. Then
\begin{equation}
f(\S_g) \geq \left(1-e^{-\frac{1}{c}}\right) f(\S^\star),
\end{equation}
where $c=\max\{c_f,1\}$ and
\begin{equation}
f(\S_g) \geq \left(1-\frac{1}{e}\right) \left(f(\S^\star)-(k-1)\epsilon_f\right).
\end{equation}
\end{proposition}

The results of Propositions \ref{lem:curv} and \ref{thm:curv} imply that if the objective function of observation selection task (see \eqref{eq:probcard}) is monotone and (weak) submodular, the greedy selection scheme that in each iteration selects an observation with the highest marginal gain satisfies the approximation bounds given in Proposition \ref{thm:curv}. Indeed, it has been shown that when the observation model is linear, D-optimality criterion is submodular \cite{krause2008near,shamaiah2010greedy}, T-optimality criterion is modular \cite{krause2008near,summers2016submodularity} while A-optimality and E-optimality measures are weak submodular \cite{bian2017guarantees,chamon2017approximate,hashemi2017randomized}.

When the model is nonlinear, which as we discussed before is frequently encountered in many applications including phase retrieval and localization, finding the posterior distribution of the unknown parameter (and hence the error covariance matrix $\M_\S$ associated with an estimator $\hat{\tt}_\S$) becomes intractable in general. 
Existing approaches mainly rely on heuristic methods to approximate the expected utility, for instance, by linearizing the model or employing Monte-Carlo methods \cite{chaloner1995bayesian,sebastiani2000maximum,flaherty2006robust,krause2008robust,davidian2017nonlinear}. For instance, in locally-optimal observation selection approach in experimental design \cite{flaherty2006robust,krause2008robust} for a nonlinear model $y_i = g_i(\tt)+\nu_i$, one linearizes the model around an initial guess $\tt_0$ (e.g. $\tt_0 = \E[\tt]$) to obtain
\begin{equation}
\begin{aligned}
\tilde{y}_i &:= y_i -g_i(\tt_0) = \nabla g_i(\tt_0)^\top\tt+\nu_i,
\end{aligned}
\end{equation}
find an approximate moment
\begin{equation}
\hat{\M}_\S =  \left(\P^{-1}+\sum_{i\in\S}\frac{1}{\sigma^2_i}\nabla g_i(\tt_0)\nabla g_i(\tt_0)^\top\right)^{-1},
\end{equation}
and use alphabetical scalarization of $\hat{\M}_\S$ as the optimality measure to select the subset $\S_g$ of observation via the greedy selection scheme. Notice that $\hat{\M}_\S$ is no longer equivalent to the error covariance (i.e. moment) matrix. Therefore optimizing scalarization of $\hat{\M}_\S$ does not necessarily result in selecting a low MSE subset.  Additionally, existing theoretical results established for the performance of the greedy algorithm hold for the approximate, linearized model and not for the original nonlinear observation model. 

In contrast to the existing observation selection methods for nonlinear models that rely on finding an approximate moment matrix, our proposed framework builds upon the idea of optimizing alphabetical scalarizations of the Van Trees' bound \cite{van2004detection} on the moments of a weakly biased estimator. As we will see in the subsequent sections, these {\it surrogate utilities} have an intuitive connection to MSE while enjoying (weak) submodularity. The Van Trees' inequality is outlined in the following theorem.
\begin{theorem}\label{thm:van}
Let $\tt$ be a collection of random unknown parameters, and let $\y_\S = \{y_i\}_{i\in\S}$ denote the collection of measurements indexed by the subset $\S$. For any estimator $\hat{\tt}_\S$ that satisfies
\begin{equation}
\int_{-\infty}^{+\infty} \nabla_\Tt\left(p_\tt(\Tt)\E_{\y|\tt}[\hat{\tt}_\S - \Tt]\right) d\Tt= \mathbf{0},
\end{equation}
it holds that
\begin{equation}\label{eq:vtb}
\M_\S\succeq \E_{\y_\S,\tt}\left[(\nabla_\Tt \log q_\tt(\Tt))(\nabla_\Tt \log q_\tt(\Tt))^\top\right]^{-1},
\end{equation}
where $q_\tt(\Tt) = p_{\tt,\y_\S}(\Tt;\y)$ is the posterior distribution of $\tt$ given $\y_\S$.
\end{theorem}
The condition stated in Theorem  \ref{thm:van} essentially quantifies to what extend the estimator is biased. Indeed, for an unbiased estimator satisfying $\E_{\y|\tt}[\hat{\tt}_\S] = \tt$, this condition is met.

The lower bound in the Van Trees' inequality cannot be computed in a closed-form for general nonlinear models. Hence, we focus our study on quadratic models since, as we show in the next section, the Van Trees' bound has a closed-form expression.
\section{Observation Selection for Quadratic Models}\label{sec:quad}
Consider the quadratic model
\begin{equation}\label{eq:quad}
y_i = \frac{1}{2} \tt^\top \X_i \tt + \z_i^\top \tt + \nu_i \; , \quad i \in [n],
\end{equation}
where $\X_i$ and $\z_i$ are known parameters (e.g., the features or the design parameters), $\nu_i\sim\N(0,\sigma^2_i)$, and $\tt$ denotes the unknown collection of parameters with prior $p_\tt(\Tt)$. We further assume the expectation is known and $\E[\tt] = \mathbf{0}$; alternatively, we can make $\tt$ centered.

Our first theoretical result, stated in Theorem \ref{thm:main1}, demonstrates that for the quadratic model in \eqref{eq:quad} and for any prior on $\tt$ with covariance matrix $\P$, the Van Trees' bound has a closed-form expression.
\begin{theorem}\label{thm:main1}
Let $\B_\S$ denote the lower bound in the Van Trees' inequality for the quadratic model \eqref{eq:quad}. Let $p_\tt(\Tt)$ be the prior on $\tt$ such that $\E[\tt] = \mathbf{0}$, and $\Cov(\tt) = \P$. Then
\begin{equation}
\B_\S = \left(\sum_{i\in\S}\frac{1}{\sigma^2_i}\left(\X_i\P\X_i^\top+\z_i\z_i^\top\right)+\I_x\right)^{-1},
\end{equation}
where 
\begin{equation}
\I_x = \E_{\tt}\left[(\nabla_\Tt \log p_{\tt}(\Tt))(\nabla_\Tt \log p_{\tt}(\Tt))^\top\right]
\end{equation}
is the Fisher information matrix associated with $p_\tt(\Tt)$.
\end{theorem}
Theorem \ref{thm:main1} opens a new avenue in the task of observation selection and information gathering for quadratic models which, as we see in our simulation results, enables selection of observations leading to lower estimation error (i.e., higher information) as compared to the approximate method based on linearization. Relying on the result of Theorem \ref{thm:main1}, we propose to use alphabetical optimality criteria (see (3) -- (6)) applied to the Van Trees' lower bound  $\B_\S$ as the utility function in the observation selection task (effectively replacing $\M_\S$ and $\P^{-1}$ with $\B_\S$ and $\I_x$, respectively). Since these utility functions aim to minimize a scalarization of Van Trees' lower bound on MSE, they have an explicit connection to MSE. We note however that similar to locally-optimal observation selection, the proposed utilities are in fact surrogates to MSE and ultimately heuristic in nature. However, due to the connection established by Theorem \ref{thm:van}, they result in selection of a more informative subset of observations compared to other heuristics. Nonetheless, the Van Trees lower bound is asymptotically tight, i.e., it is tight in the high signal-to-noise ratio settings or in the case of sufficiently large number of observations. Hence, we expect to select a near-optimal subset by using the proposed selection criteria in such settings.

Compared to existing robust methods, \cite{flaherty2006robust} consider robustness with respect to the Jacobian of the unknown parameters in the linearized model and apply semi-definite programming with E-optimality criterion. \cite{krause2008robust} consider robustness with respect to the initial guess. They do so by performing linearization over multiple starting point and use the saturated algorithm for selection. In large-scale problems one needs to consider many starting points which is impractical. Since for quadratic models we can efficiently find the Van Trees' inequality without any approximation, it is meaningful to utilize such structural property. However, for general nonlinear models, robust methods such as \cite{flaherty2006robust,krause2008robust} are indeed advantageous.

The question that remains to be answered is whether the greedy observation selection method that provides a provably near-optimal selection for the linear models still enjoys similar theoretical performance guarantees. To this end, we demonstrate such theoretical performance guarantees for the greedy algorithm by showing weak submodularity of the proposed optimality criteria.
\section{Near-Optimal Greedy Observation Selection}\label{sec:alg}
In the following theorems we consider alphabetical scalarizations of the Van Trees' bound $\B_S$ defined in Theorem \ref{thm:main1} and show that they are monotonically increasing as well as either modular, submodular, or weak submodular. These results illustrate not only that the proposed optimality criteria deal with the quadratic model without resorting to any approximations, but also that one can use the greedy observation selection method of Algorithm 1 to find a near-optimal subset of observations with performance guarantees established in Proposition \ref{thm:curv}. Proofs of the subsequent results are established by employing tools from linear algebra and matrix analysis such as Weyl's inequality, Sylvester's determinant identity, matrix inversion lemma, and Courant–Fischer min-max theorem \cite{bellman1997introduction}.
\begin{theorem}
	Instate the notation and hypothesis of Theorem \ref{thm:main1}. The T-optimality of the Van Trees' bound, i.e.,
	\begin{equation}
	f^T(\S) = \mathrm{Tr}\left(\B_\S^{-1}\right) - \mathrm{Tr}\left(\I_x\right),
	\end{equation}
	is monotone and modular.
\end{theorem}
\begin{theorem}
	Instate the notation and hypothesis of Theorem \ref{thm:main1}. The D-optimality of the Van Trees' bound, i.e.,
	\begin{equation}
	f^D(\S) = \log\det\left(\B_\S^{-1}\right) - \log\det\left(\I_x\right),
	\end{equation}
	is monotone and submodular.
\end{theorem}
\begin{theorem}
	Instate the notation and hypothesis of Theorem \ref{thm:main1}. The E-optimality of the Van Trees' bound, i.e.,
	\begin{equation} 
	f^E(\S) = \lambda_{\min}\left(\B_\S^{-1}\right) -\lambda_{\min}\left(\I_x\right), 
	\end{equation}
	is monotone and weak submodular. Further, its additive and multiplicative weak-submodularity constants satisfy
	\begin{equation}
	\begin{aligned}
	c_{f^E}&\le 
	\max_{j \in \Xs} \frac{\lambda_{\max}\left(\I_j\right)}{\lambda_{\min}\left(\I_j\right)}, \\ \epsilon_{f^E}&\le 
	\max_{j \in \Xs} \left(\lambda_{\max}\left(\I_j\right) - \lambda_{\min}\left(\I_j\right)\right),
	\end{aligned}
	\end{equation}
	where $\I_j = \frac{1}{\sigma^2_j}\left(\X_j\P\X_j^\top+\z_j\z_j^\top\right)$.
\end{theorem}
The term $\I_j$ is reflective of the amount of {\it information} captured by the $j\ts{th}$ observation. In this regard, Theorem 5 states that if the difference between the minimum and maximum information of individual observations is small, the E-optimality of $\B_\S$ is nearly submodular. Hence, the greedy observation selection scheme is expected to find a good (informative) subset.
\begin{theorem}
	Instate the notation and hypothesis of Theorem \ref{thm:main1}. The A-optimality of the Van Trees' bound, i.e.,
	\begin{equation}
	f^A(\S) = \mathrm{Tr}\left(\I_x^{-1}\right)-\mathrm{Tr}\left(\B_\S\right), 
	\end{equation}
	is monotone and  weak submodular. Furthermore, if $\z_i = \mathbf{0}$ and $\X_i= \x_i\x_i^\top$, the additive and multiplicative weak-submodularity constants satisfy
	\begin{equation}\label{eq:aoptcon}
	c_{f^A}\leq \max_j \gamma_j, \quad \epsilon_{f^A}\leq \max_j \frac{\lambda_{\min}(\B_{[n]})^{2}}{\lambda_{\max}(\sigma_j^2\P+\I_x^{-1})}(\gamma_j-1)
	\end{equation}   
	where
	\begin{equation}
	\gamma_j = \frac{\lambda_{\max}(\I_x^{-1})^{2}(\lambda_{\max}(\sigma_j^2\P)+1)}{\lambda_{\min}(\B_{[n]})^{2}(\lambda_{\min}(\sigma_j^2\P)+1)}.
	\end{equation}
\end{theorem}
The conditions on additive and multiplicative WSCs of A-optimality in Theorem 6 are essentially conditions on observations' SNR. Specifically, $\gamma_j$ can be interpreted as the normalized SNR of the $i\ts{th}$ observation. In this regard, according to  \eqref{eq:aoptcon} the behavior of the A-optimality approaches that of a submodular function if the energy of observation with highest SNR is relatively small. Additionally, if observations are fairly uncorrelated, conditions in \eqref{eq:aoptcon} are met even if the SNRs are large.
Furthermore, we note that the simplifying condition $\z_i = \mathbf{0}$ and $\X_i= \x_i\x_i^\top$ is motivated by the phase retrieval application, studied in Section \ref{sec:sim}.
\begin{figure*}[t]
	\centering
	\minipage[t]{1\textwidth}
	\begin{subfigure}[t]{.3333\textwidth}
		\includegraphics[width=\textwidth]{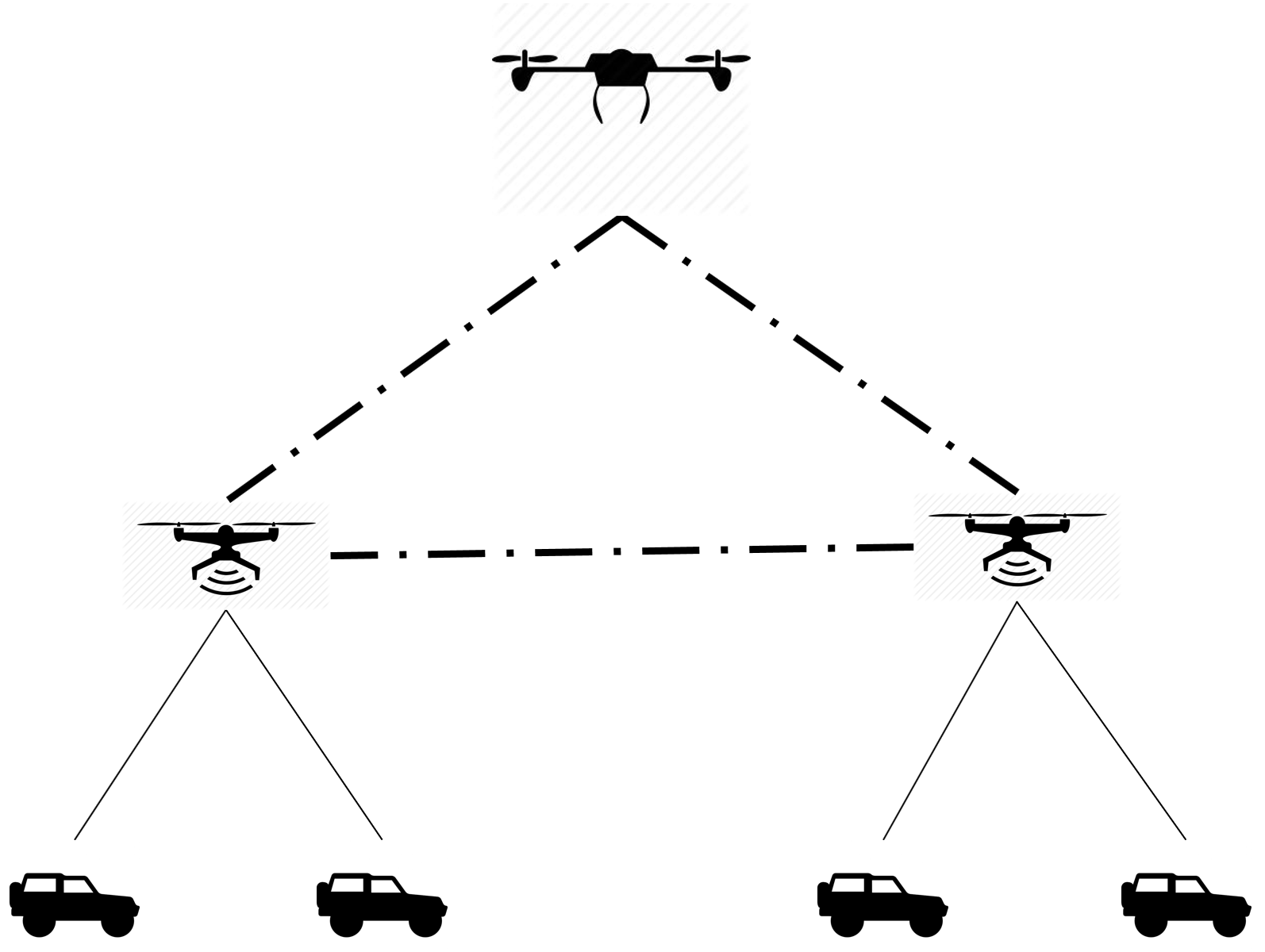}
		\caption{\footnotesize  tracking via a swarm of UAVs}
	\end{subfigure}
	\begin{subfigure}[t]{.3333\textwidth}
		\includegraphics[width=\textwidth]{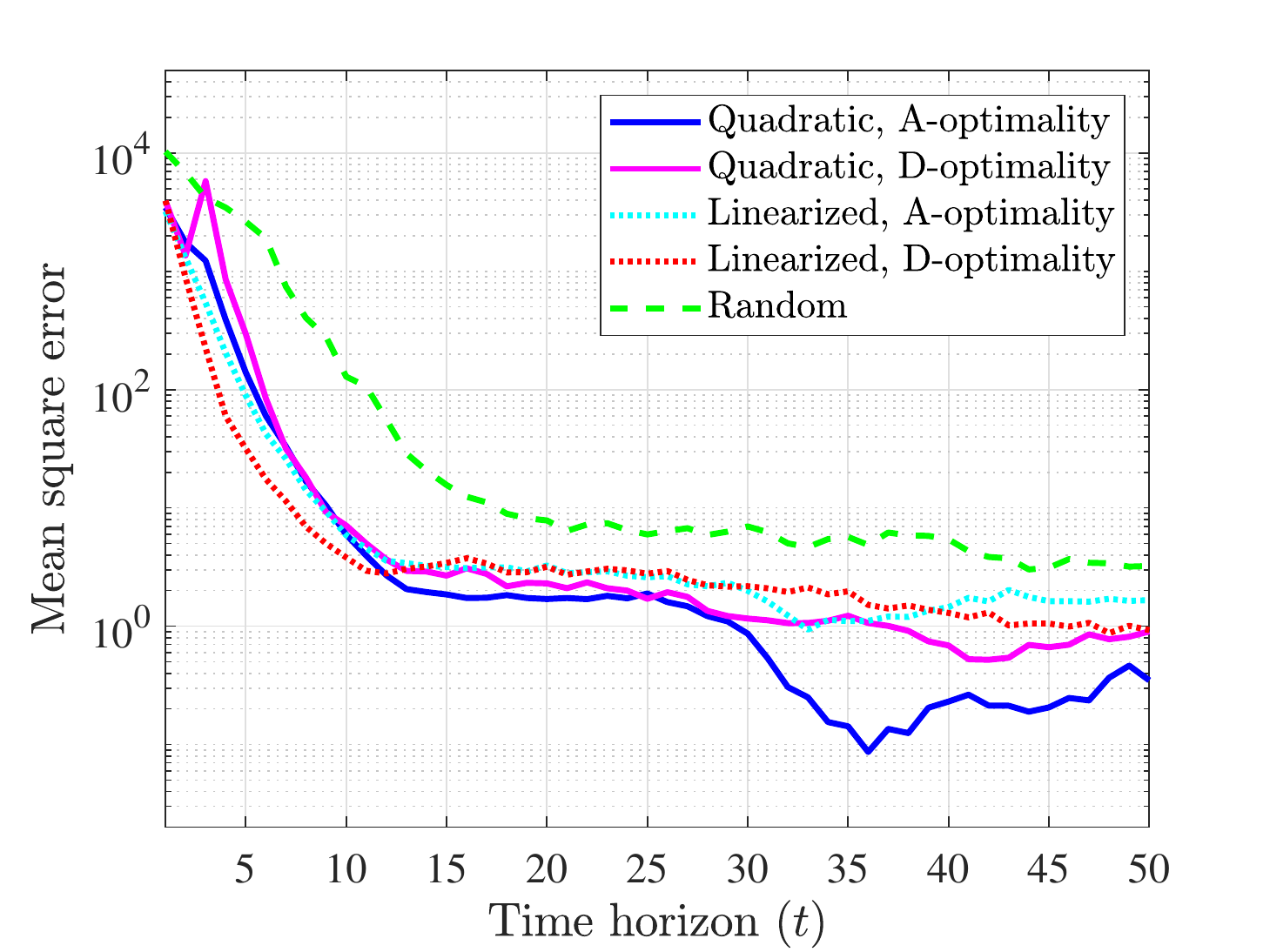}
		\caption{\footnotesize   tracking under identical noise powers}
	\end{subfigure}
	\begin{subfigure}[t]{.3333\textwidth}
		\includegraphics[width=\textwidth]{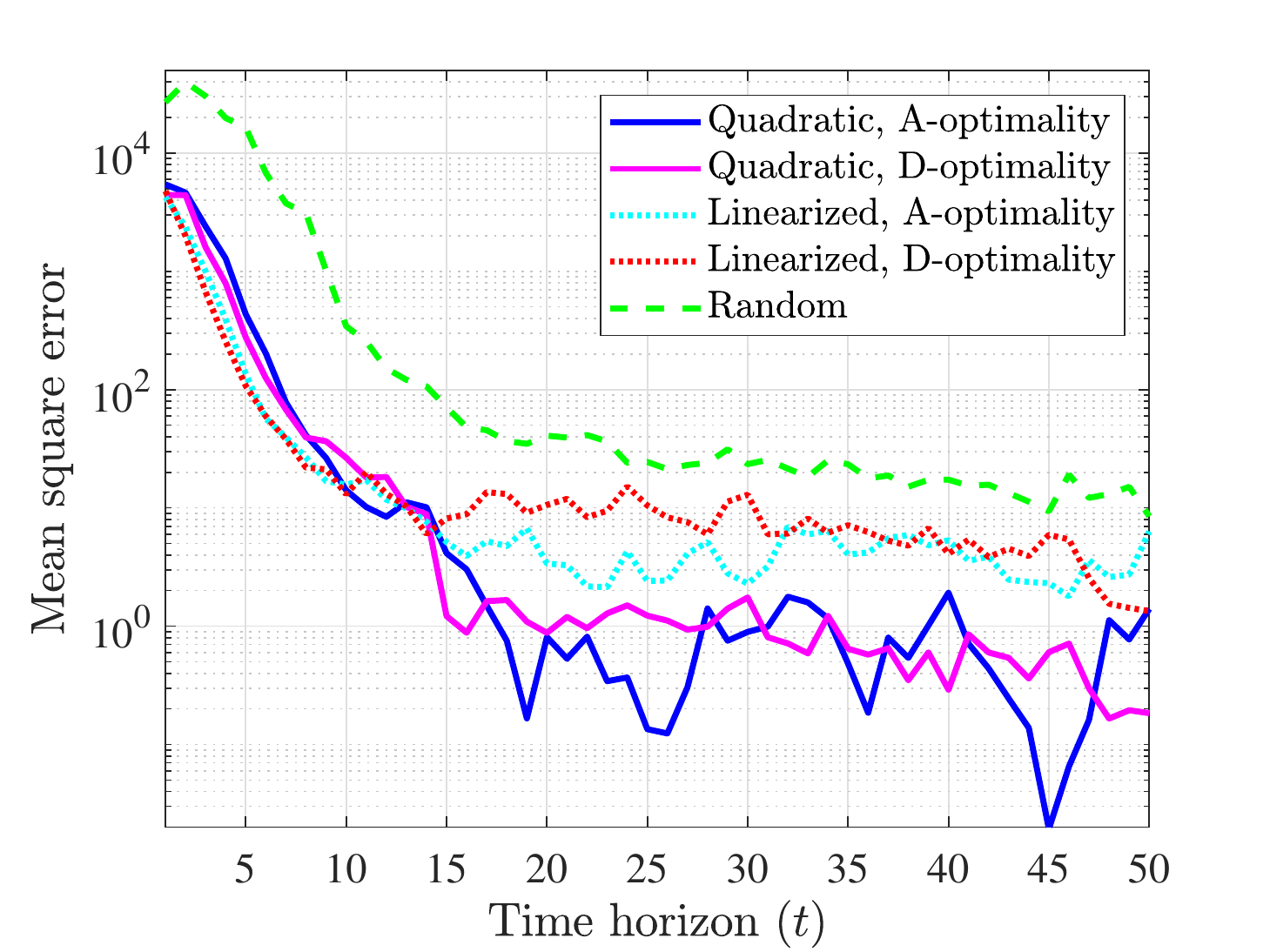}
		\caption{\footnotesize  tracking under random noise powers}
	\end{subfigure}
	\caption{Comparison of MSEs for random, linearized, and quadratic observation selection schemes in the multi-target tracking application.}
	\label{fig:uavs}
	\endminipage 
\end{figure*}
\section{Applications and Numerical Tests}\label{sec:sim}
In this section we test the efficacy of the proposed quadratic observation selection optimality criteria in two applications: multi-object tracking using radar measurements and phase retrieval from magnitude measurements. We limit our experiments to A-optimality and D-optimality for their intuitive interpretation: A-optimality of $\B_\S$ for a minimum variance unbiased estimator attaining \eqref{eq:vtb} with equality is equivalent to the mean-square error (MSE), while, in the context of parameter estimation in linear models with independent and homoscedastic noise, D-optimality is equivalent to the maximization of entropy of parameters \cite{krause2008near}. 
\subsection{Constrained multi-target tracking}
\begin{figure*}[t]
	\centering
	\minipage[t]{1\textwidth}
	\begin{subfigure}[t]{.3333\textwidth}
		\includegraphics[width=\textwidth]{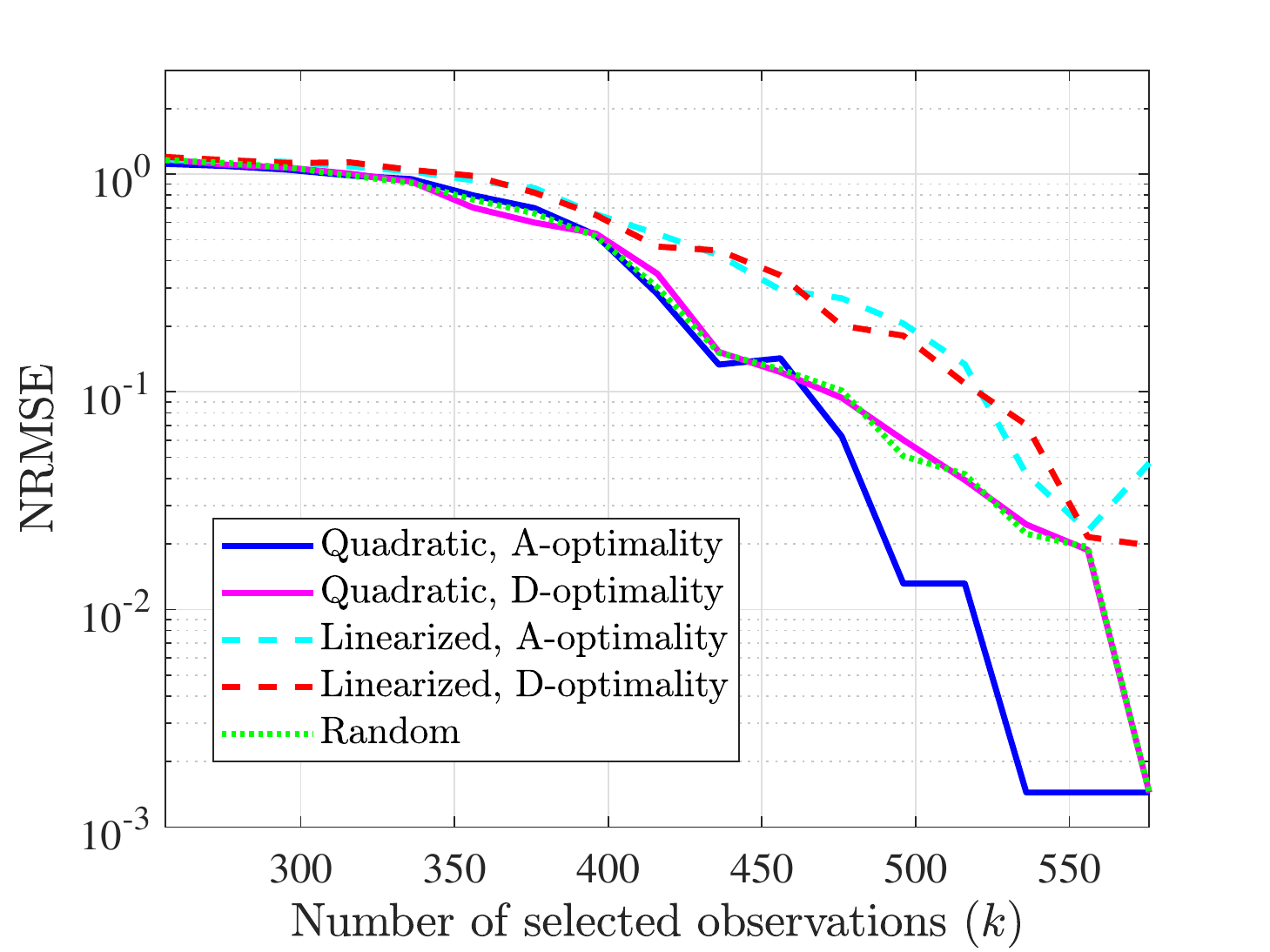}
		\caption{\footnotesize  Gaussian with identical noise powers}
	\end{subfigure}
	\begin{subfigure}[t]{.3333\textwidth}
		\includegraphics[width=\textwidth]{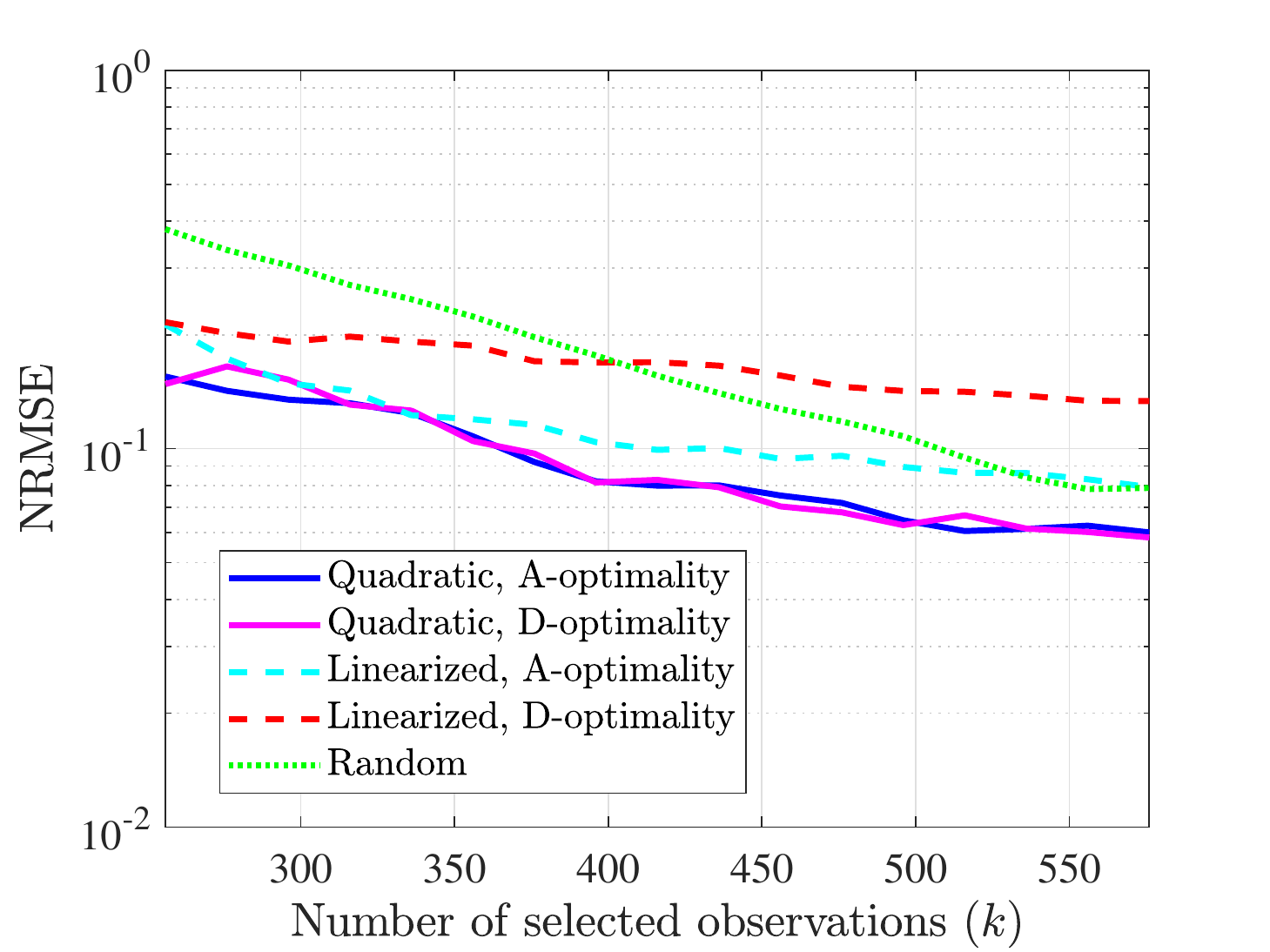}
		\caption{\footnotesize  DFT with identical noise powers}
	\end{subfigure}
	\begin{subfigure}[t]{.3333\textwidth}
		\includegraphics[width=\textwidth]{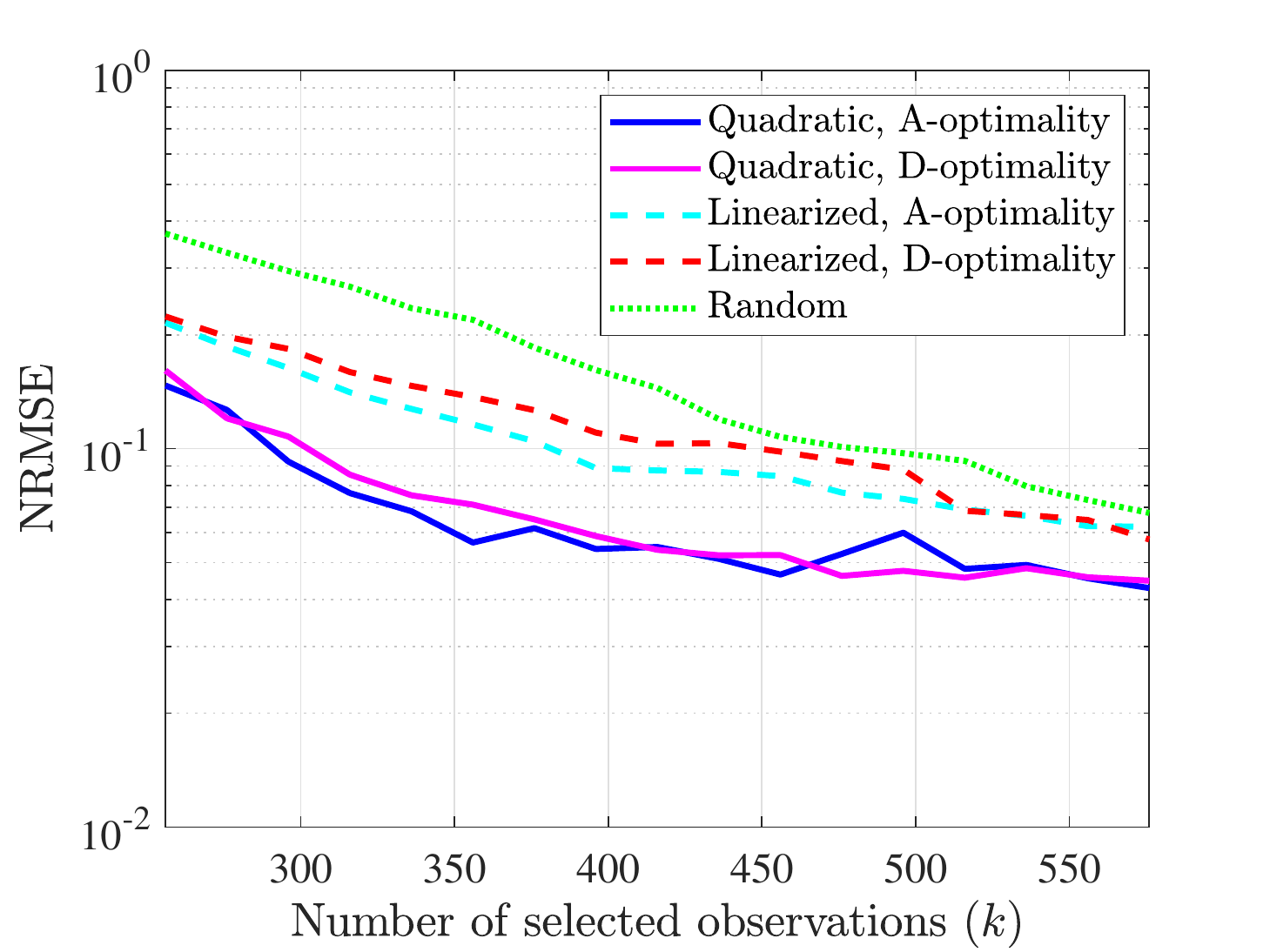}
		\caption{\footnotesize DFT with random noise powers}
	\end{subfigure}
	\caption{Comparison of NRMSEs for random, linearized, and quadratic observation selection schemes for the phase retrieval.}
	\label{fig:phase1}
	\endminipage 
\end{figure*}
We study a multi-object tracking application where a control unit surveys an area via a swarm of unmanned aerial vehicles (UAVs) (Figure \ref{fig:uavs}(a)). The UAVs are equipped with GPS and radar systems, and can communicate with each other over locally established communication channels. Due to limitations on the rate of communication between the {\it swarm leaders} and the control unit and in order to reduce delays in tracking due to intensive computation, only a subset of the gathered measurements is communicated to the control unit. In order to track the locations of the objects in the environment, the control unit employs extended Kalman filter (EKF) to process the received measurements. Therefore, the goal of the swarm leaders is to perform {\it observation selection and information gathering} and select a subset of range and angular measurements such that (i) the communication constraint is satisfied, and (ii) the mean-square error of the EKF estimate of the objects' locations is minimized. Let $\mathbf{u}_i^{t}$ and $\mathbf{s}_j^{t}$ denote the location of the $i\ts{th}$ UAV and the $j\ts{th}$ object, respectively.\footnote{For simplicity, objects and UAVs' altitudes are fixed.} 
The range measurements of the radar system follow a quadratic model:
\begin{equation}\label{eq:range}
r_{ij} = \frac{1}{2}\|\mathbf{u}_i^{t}-\mathbf{s}_j^{t}\|_2^2+\nu_{ij}.
\end{equation}

A comparison with \eqref{eq:quad} reveals that, in this model, (relative to each pair of UAV-target) $\X_i$'s are all equal to the identity matrix and $\z_i = \mathbf{u}_i^{t}-\hat{\mathbf{s}}_j^{t-1}$ where $\hat{\mathbf{s}}_j^{t-1}$ is the prior estimates of objects' locations (after centering $\mathbf{s}_j^{t}$ around its approximate expectation $\E[\mathbf{s}_j^{t}] \approx \hat{\mathbf{s}}_j^{t-1}$).
Therefore, we can employ the proposed quadratic observation selection framework directly. The angular measurements on the other hand have the nonlinear form
\begin{equation}\label{eq:angular}
\alpha_{ij} = \text{arctan}\frac{{u}_i^{t}(1)-{s}_j^{t}(1)}{{u}_i^{t}(2)-{s}_j^{t}(2)}+\eta_{ij}.
\end{equation}
To select the angular measurements, we follow the locally optimal approach of \cite{flaherty2006robust} and linearize \eqref{eq:angular} around the prior estimates of objects' locations $\hat{\mathbf{s}}_j^{t-1}$.

We implement a Monte Carlo simulation with 50 independent instances where 10 moving objects are initially uniformly distributed in a $5\times 10$ area. At each time instance, the objects move in a random direction with a constant velocity set to 0.2. The swarm consists of $10$ UAVs, equidistantly spread over the area, that move according to a periodic \textit{parallel-path} search pattern \cite{vincent2004framework}. 
The initial phases of the UAVs' motions are uniformly distributed to provide a better coverage of the area. The UAVs can acquire range and angular measurements of the objects that are within the maximum radar detection range. The maximum radar detection range is set such that, at each time step, the UAVs together collect approximately 130-170 range and angular measurements. The communication bandwidth constraints limit the number of measurements
transmitted to the control unit to $10\%$ of the gathered measurements.
We consider two noise models: in the first
scenario the noise terms $\nu_{ij}, i=1,\ldots,10, j=1,\ldots,10$ are i.i.d. Gaussian noises with $\sigma_{ij} = 0.01$ 
while in the second scenario we logarithmically space the
interval $(0:001;0:01)$ to generate 10 points and select $\sigma_{ij}$ for
each observation uniformly at random from one of these 10 numbers. We use A-optimality and D-optimality as the selection criteria and assess the performance of different schemes using the MSE of the EKF estimates of objects' locations.

Figure \ref{fig:uavs}(b) and Figure \ref{fig:uavs}(c) illustrate the results for the two noise models. At the beginning of tracking, all schemes have relatively high error. However, since the observations selected by the proposed schemes are chosen according to the exact range model,
as time passes the MSE of the proposed scheme becomes significantly lower than those of locally optimal and random selection methods (especially under the A-optimality criterion). Figure \ref{fig:uavs} also depicts that the MSE of the estimates formed from the observations selected by the proposed quadratic observation selection scheme using A-optimality is lower than the MSE achieved by selecting the observation via D-optimality. This reduction is due to the fact that if the estimator (here the EKF) is a minimum variance unbiased estimator attaining \eqref{eq:vtb} with equality, the A-optimality scalarization of the Van Trees' bound becomes equivalent to the MSE, the performance measure shown in Figure \ref{fig:uavs}. Therefore, intuitively one expects to achieve lower MSE using the A-optimality scalarization of the Van Trees' bound, which is the case in this experiment.

\subsection{Phase retrieval from magnitude measurements}
Phase retrieval is the task of predicting a possibly complex unknown variable from its magnitude measurements. Specifically, we are given measurements \cite{candes2015phase,candes2015phasea}
\begin{equation}\label{eq:pm}
    y_i = g_i(\tt) = \frac{1}{2}|\a_i^\ast\tt|^2+\nu_i,
\end{equation}
where $\a_i \in \mathbb{C}^{n}$ is the $i\ts{th}$ measurement vectors, and $\ast$ denotes the conjugate transpose operator. The model in \eqref{eq:pm} is an instance of the quadratic model \eqref{eq:quad} as it can be written as $y_i = \frac{1}{2}\tt^\ast(\a_i\a_i^\ast)\tt+\nu_i$. However, since $g_i(\tt)$ is not holomorphic (i.e. complex differentiable) in general, the results of Theorem \ref{thm:main1} cannot be applied directly. Therefore, we approximate the Van Trees' bound by using the partial Wirtinger derivative of $g_i(\x)$. That is, we treat $\tt$ as a real-valued variable and compute the gradient of $g_i(\tt)$ to obtain 
\begin{equation}\label{eq:pn}
\begin{aligned}
\hat{\B}_\S &= \left(\sum_{i\in\S}\frac{1}{\sigma^2_i}\a_i\a_i^\ast\P\a_i\a_i^\ast+\I_x\right)^{-1}
\\&= \left(\sum_{i\in\S}\frac{1}{\tilde{\sigma}^2_i}\a_i\a_i^\ast+\I_x\right)^{-1},
\end{aligned}
\end{equation}
where $\tilde{\sigma}^2_i = {\sigma^2_i}\slash({\a_i^\ast\P\a_i})$. The expression for $\hat{\B}_\S$ in \eqref{eq:pn} reveals an interesting {\it noise adjustment} effect. Because of the specific structure of the quadratic model in phase retrieval, i.e., $\X_i$'s are of rank 1, the (approximate) Van Trees' bound resembles the structure of the moment matrix in linear models except that the noise powers are now observation-specific and are related to $\tt$ through the covariance matrix $\P$.\looseness=-1 

In order to investigate the performance of the proposed observation selection framework, we consider the task of predicting a complex signal $\tt\in \mathbb{C}^{n}$ with $n = 128$ by selecting a subset of size $k$ from $m = 1280$ observations. The complex signal $\tt$ is distributed as standard circularly-symmetric complex Gaussian, and we vary $k$ from $256$ to $576$. We consider two standard scenarios where for each we average the results over 50 independent instances, and use the Wirtinger flow algorithm \cite{candes2015phasea} as the oracle estimator.
\begin{figure*}[t]
	\centering
	\minipage[t]{1\textwidth}
	\begin{subfigure}[t]{.3333\textwidth}
		\includegraphics[width=\textwidth]{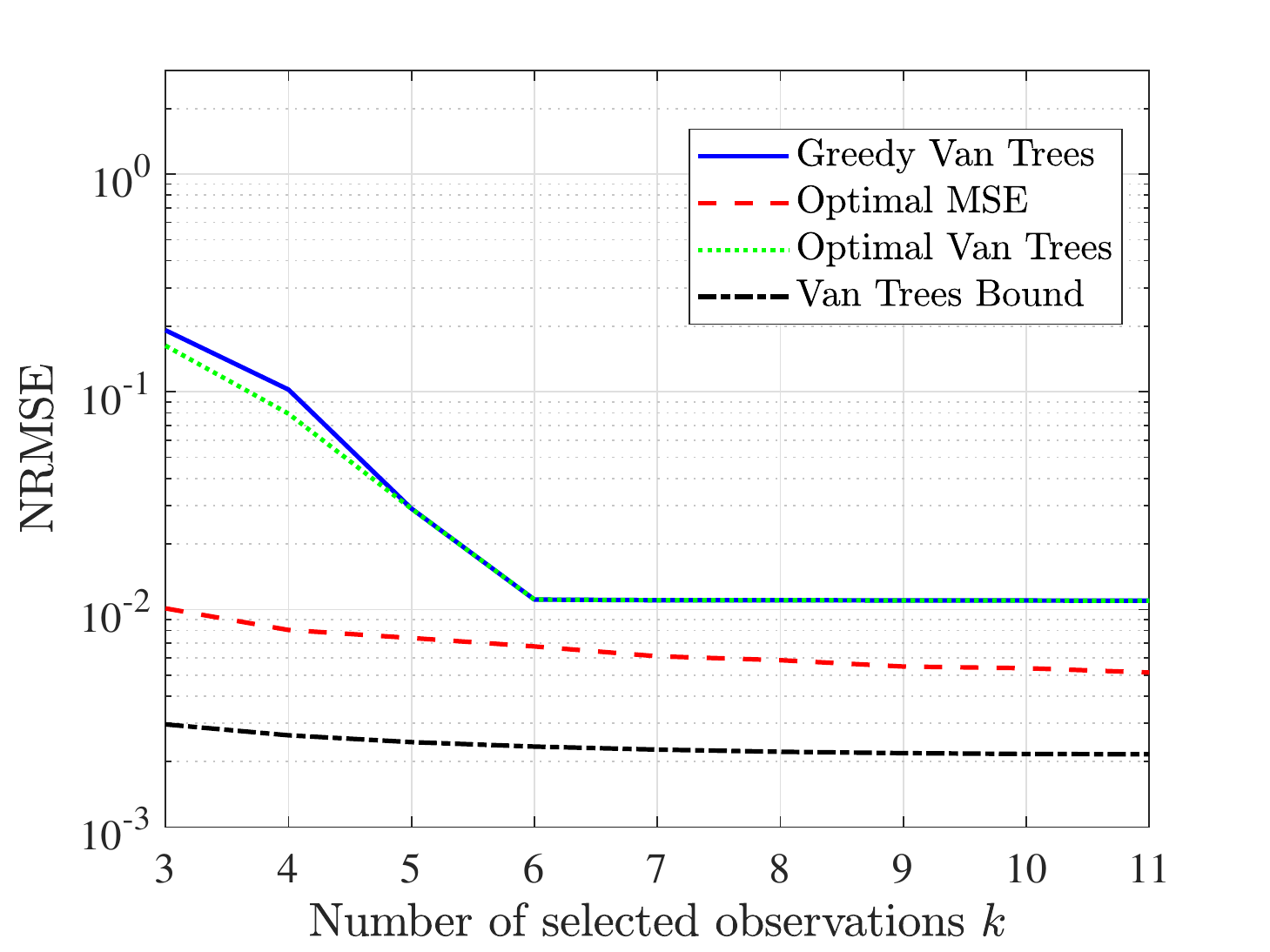}
		\caption{\footnotesize  MSE vs Van Trees' bound}
	\end{subfigure}
	\begin{subfigure}[t]{.3333\textwidth}
		\includegraphics[width=\textwidth]{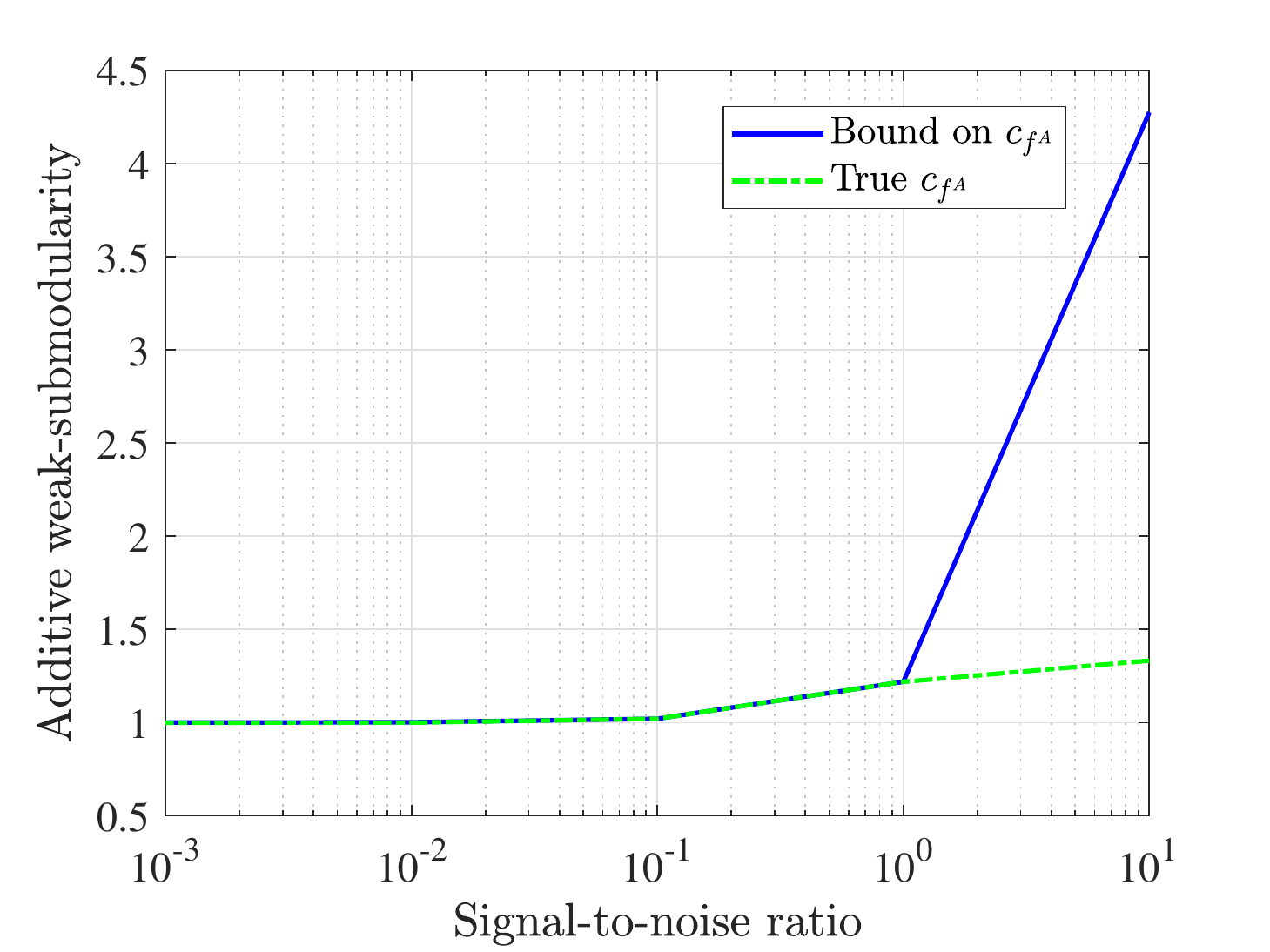}
		\caption{\footnotesize  additive WSC}
	\end{subfigure}
	\begin{subfigure}[t]{.3333\textwidth}
		\includegraphics[width=\textwidth]{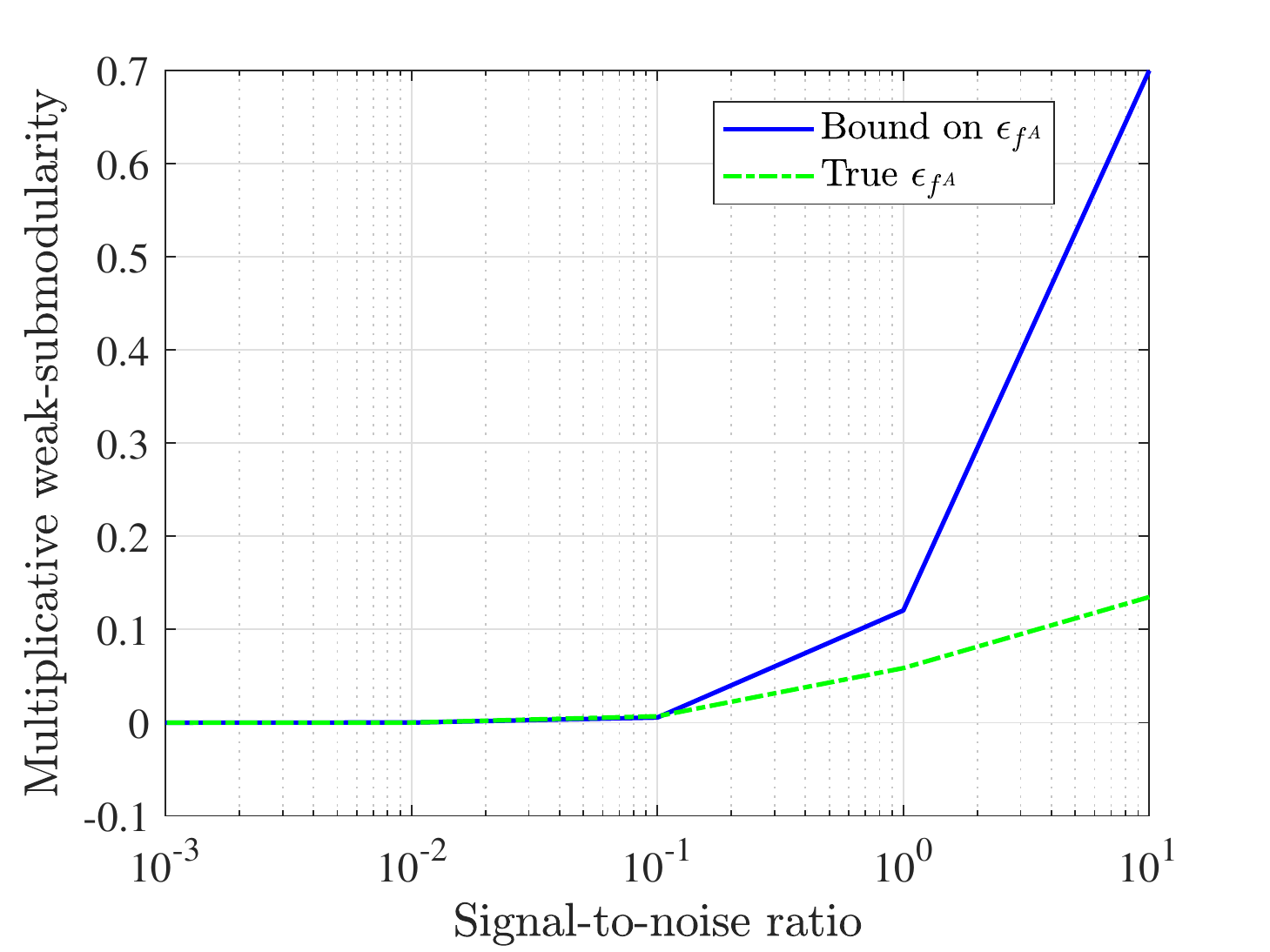}
		\caption{\footnotesize multiplicative WSC}
	\end{subfigure}
	\caption{Evaluation of theoretical results in Section 4 for a small-scale  phase retrieval task.}
	\label{fig:thm}
	\endminipage 
\end{figure*}
\subsubsection{Complex Gaussian measurements} 
First we examine the case in which all the elements in the measurement vectors $\a_i$ are independent random variables following a standard complex Gaussian distribution. Thus every element is regarded as a random variable in which the real part and the imaginary part are drawn from the standard Gaussian distribution $\mathcal{N}(0,1)$ independently. We also assume that the noise powers are $\sigma_i = 0.001$ for all measurements. 

Figure \ref{fig:phase1} (a) illustrates the normalized root MSE (NRMSE) results. The subset selected to achieve the A-optimality of $\hat{\B}_\S$ provides the lowest estimation error. This observation is again supported by the relation of A-optimality and MSE for minimum variance unbiased estimators satisfying \eqref{eq:vtb} with equality. The performance of subset selected to achieve the D-optimality of $\hat{\B}_\S$ and that of random subset selection are virtually the same, which is also expected since D-optimality is intuitively achieved by selecting observations according to the entropy gains. Since $\a_i$'s are i.i.d. and the noise powers are equal for all measurements, the entropy gain is similar for all measurements. Hence, D-optimality acts relatively similarly to random subset selection in this scenario. Locally optimal observation selection schemes achieve the lowest performance, partly because of the sensitivity of these schemes to the initial estimate used in linearization. 
\subsubsection{Discrete Fourier measurements} 
In traditional phase retrieval problems, the measurements are the magnitude of the Fourier transform of the signal. Hence, next we consider the $\a_i$'s to be rows of an $m\times m$ DFT matrix restricted to the first $n$ columns. Figure \ref{fig:phase1} (b) shows the NRMSE results, where, as we see, the proposed quadratic observation selection framework achieves the lowest error. Note that, in contrast to the case of complex Gaussian measurement vectors, the entropy gain of the measurements is expected to be different under this scenario. Therefore, D-optimality of $\hat{\B}_\S$ results in a superior performance compared to the random observation selection scheme. Furthermore, since selecting more observations reduces the estimation error, as $k$ increases, the performance gap between different schemes decreases. 

Finally, we consider a case where the noise powers are not identical for all measurements. Rather, we logarithmically space the interval $(0.001,0.01)$ to generate $10$ points and select $\sigma_i$ for each observation uniformly at randomly from one of these $10$ numbers. Figure \ref{fig:phase1} (c) depicts the NRMSE results, where we again observe the superiority of the proposed optimality criteria to select a subset of observation with the lowest estimation error. 
\subsection{Evaluation of theoretical results}
To numerically study the connection between Van Trees' bound and MSE, we consider a phase retrieval problem with $m = 12$ Gaussian observations where we find the optimal subsets via exhaustive search. Figure \ref{fig:thm} (a) shows comparison of NRMSE of greedy maximization of A-optimality of Van Trees’ bound, maximization of A-optimality of Van Trees’ bound via exhaustive search, and maximization of MSE via exhaustive search. We also plot the theoretical Van Trees’ bound (see (20)) evaluated at the set found by the approach of maximization of MSE via exhaustive search. As we see, our approach performs similarly to the exhaustive Van Trees’ bound optimization which shows near optimality of the greedy selection scheme. Also, as we select more observations all three selection schemes approach the theoretical Van Trees’ bound, which is supported by the asymptotic exactness of Van Trees’ bound. 

Figure \ref{fig:thm}(b) and Figure \ref{fig:thm}(c) shows the true values of additive and multiplicative WSCs found by exhaustive search as well as our established bounds for A-optimality of Van Trees’ bound in Theorem 6. As we see, with smaller SNRs, the gap between the two is negligible however for larger SNRs the established bounds are relatively weak.

\section{Conclusion}
We studied the task of observation selection and information gathering for quadratic measurement models.  Since the moment matrix of the optimal estimator is unknown due to nonlinearity of the observation model, typical optimality criteria no longer possess connections to MSE, the desired performance metric. To address this challenge, we derived new criteria by relying on the Van Trees' inequality and proved that they are monotone (weak) submodular set functions under certain conditions on the unknown parameters and the features of the model. Following these results, we developed an efficient greedy observation selection algorithm with theoretical bounds on its achievable utility.
\section*{Acknowledgements}
This work was supported in part by NSF 1809327, DARPA grant D19AP00004, and ONR grants N00014-18-1-2829 and N00014-19-1-2054. We thank Alexandros G. Dimakis and Ethan R. Elenberg for their constructive discussion and also thank the reviewers for their valuable feedback.
\bibliographystyle{apalike}

\section*{Supplementary Material}
\section*{Proof of Proposition 1}
First note that we can define $c_f$ equivalently as $c_f=\max_{l=1}^{n-1}{{\cal C}_l}$ where 
\begin{equation}\label{eq:cel}
{\cal C}_l=\max_{(\S,\T,i)\in \Xs_l}{f_i(\T)\slash f_i(\S)},
\end{equation}
and $\Xs_l = \{(\S,\T,i)|\S \subseteq \T \subset \Xs, i\in \Xs \backslash \T, |\T\backslash\S|=l\}$.
Now, let $\S \subset \T$ and $\T \backslash \S=\{j_1,\dots,j_r\}$. Then,
\begin{multline}
f(\T)-f(\S)=f(\S\cup \{j_1,\dots,j_r\})-f(\S)\\\qquad\quad={f_{j_1}(\S)}+{f_{j_2}(\S\cup\{j_1\})}+\dots\\+{f_{j_r}(\S\cup\{j_1,\dots,j_{r-1}\})}.
\end{multline}
Applying \eqref{eq:cel} yields  
\begin{equation}\label{ds}
\begin{aligned}
f(\T)-f(\S)&\leq {f_{j_1}(\S)}+{\cal C}_1{f_{j_2}(\S)}+\dots+{\cal C}_{r-1}{f_{j_r}(\S)}\\
&={f_{j_1}(\S)}+\sum_{l=1}^{r-1}{\cal C}_l{f_{j_t}(\S)}.
\end{aligned}
\end{equation}
Note that \eqref{ds} is invariant to the ordering of elements in $\T \backslash \S$. In fact, it is straightforward to see that given ordering $\{j_1,\dots,j_r\}$, one can choose a set $\mathcal{Q}=\{{\cal P}_1,\dots,{\cal P}_r\}$ with $r$ permutations -- e.g., by defining the right circular-shift operator ${\cal P}_t(\{j_1,\dots,j_r\})=\{j_{r-t+1},\dots,j_1,\dots\}$ for $1 \leq t\leq r$ -- such that ${\cal P}_p(j)\neq{\cal P}_q(j)$ for $p\neq q$ and $\forall j\in \T \backslash \S$. Hence, \eqref{ds} holds for $r$ such permutations. Summing all of these $r$ inequalities we obtain
\begin{equation}\label{qs}
\begin{aligned}
f(\T)-f(\S)&\leq \frac{1}{r}\left(1+\sum_{l=1}^{r-1}{\cal C}_l\right)\sum_{j\in \T \backslash \S}{f_{j}(\S)}\\&\leq  \frac{1}{r}\left(1+(r-1)c_f\right)\sum_{j\in \T \backslash \S}{f_{j}(\S)}.
\end{aligned}
\end{equation}

Next, we prove the second inequality. Note that we can define $\epsilon_f =\max_{l=1}^{n-1}\epsilon_l$ where $\epsilon_l = \max_{(\S,\T,i)\in \Xs_l}{f_i(\T)- f_i(\S)}$. Using a similar argument as the one that we used for $c_f$, for any $\S \subset \T$ and $\T \backslash \S=\{j_1,\dots,j_r\}$, it holds that
\begin{equation}\label{qs1}
\begin{aligned}
f(\T)-f(\S)&\leq \sum_{l=1}^{r-1}\epsilon_l+\sum_{j\in \T \backslash \S}{f_{j}(\S)}\\&\leq (r-1)\epsilon_f+\sum_{j\in \T \backslash \S}{f_{j}(\S)},
\end{aligned}
\end{equation}
which completes the proof.
\section*{Proof of Proposition 2}
The proof follows the classical proof of greedy maximization of submodular functions given in \cite{nemhauser1978analysis}. We first prove the performance bound stated in terms of $c_f$. Consider $\S_i$, the set generated at the end of the $i\ts{th}$ iteration of the greedy algorithm and assume $|\S^\star\backslash S_t^{(i)}|=r\leq k$.
Employing Proposition 1 with $\S=\S_i$ and $\T=\S^\star\cup \S_i$, and using monotonicity of $f$ yields
\begin{equation}
\begin{aligned}
\frac{f(\S^\star)-f(\S_i)}{\frac{1}{r}\left(1+(r-1)c_f\right)}&\leq \frac{f(\S^\star\cup \S_i)-f(\S_i)}{\frac{1}{r}\left(1+(r-1)c_f\right)}\\&\leq  \sum_{j\in\S^\star\backslash \S_i}f_j(\S_i)\\&\leq r (f(\S_{i+1})-f(\S_i)),
\end{aligned}
\end{equation}
where we use the fact that the greedy algorithm selects the element with the maximum marginal gain in each iteration.
It is easy to verify, e.g., by taking the derivative, that $\frac{1}{r}\left(1+(r-1)c_f\right)$ is decreasing (increasing) with respect to $r$ if $c_f<1$ ($c_f>1$). Let $c=\max\{c_f,1\}$. Then $\frac{1}{r}(1+(r-1){\cal C}_{\max}) \leq c$. Therefore, using the fact that $r\leq k$ we get
\begin{equation}
f(\S^\star)-f(\S_i) \leq ck (f(\S_{i+1})-f(\S_i)).
\end{equation}
By induction and due to the fact that $f(\emptyset) = 0$ we obtain
\begin{equation}
f(\S_g)\geq \left(1-\left(1-\frac{1}{kc}\right)^k\right)f(\S^\star) \geq \left(1-e^{-\frac{1}{c}}\right) f(\S^\star),
\end{equation}
where we use  the fact that $(1+x)^y\leq e^{xy}$ for $y>0$.
The proof of second inequality is almost identical except we employ the second result of Proposition 1 to begin the proof.
\section*{Centering $\tt$ in Quadratic Models}
In (19), defining $\tilde{\tt} = \tt - \E[\tt]$ yields
\begin{equation}
\begin{aligned}
y_i &= \frac{1}{2} (\tilde{\tt}+\E[\tt])^\top \X_i (\tilde{\tt}+\E[\tt]) + \z_i^\top (\tilde{\tt}+\E[\tt]) + \nu_i\\
&=\frac{1}{2} \tilde{\tt}^\top \X_i \tilde{\tt} +\frac{1}{2}(\X_i\E[\tt]+\X_i^\top\E[\tt]+2\z_i)^\top \tilde{\tt}\\&\qquad+\frac{1}{2}\E[\tt]^\top\X_i\E[\tt]+\nu_i.
\end{aligned}
\end{equation}
Thus, we obtain a new quadratic model $\tilde{y}_i = \frac{1}{2} \tilde{\tt}^\top \X_i \tilde{\tt} + \tilde{\z_i}^\top \tilde{\tt} + \nu_i$ with zero-mean unknown parameters $\tilde{\tt}$, where $\tilde{y}_i = y_i-\frac{1}{2}\E[\tt]^\top\X_i\E[\tt]$, and 
$\tilde{\z_i} = \frac{1}{2}(\X_i\E[\tt]+\X_i^\top\E[\tt]+2\z_i)$.
\section*{Proof of Theorem 2}
Let $q_\tt(\Tt) = p_{\tt,\y_\S}(\Tt;\y)$ denote the posterior distribution of $\tt$ given $\y_\S$, $\G_\S = \mathrm{diag}(\{\sigma^2_i\}_{i\in\S})$ denote the noise covariance matrix $\Cov(\v_\S)$, and define 
\[\uu_\S = \mathrm{vec}(\{\frac{1}{2} \tt^\top \X_i \tt + \z_i^\top \tt\}_{i\in\S}).\]
Then, the Van Trees' bound is found as
\begin{equation}\label{eq:bs1}
\begin{aligned}
\B_\S^{-1} & = \E_{\y_\S,\tt}[(\nabla_\Tt \log q_\tt(\Tt))(\nabla_\Tt \log q_\tt(\Tt))^\top]\\
& = \E_{\y_\S,\tt}[(\nabla_\Tt \log p_{\y_\S|\tt}(\y;\Tt)p_\tt(\Tt))\\&\qquad\qquad\qquad(\nabla_\Tt \log p_{\y_\S|\tt}(\y;\Tt)p_\tt(\Tt))^\top]\\
& = \E_{\y_\S,\tt}[(\nabla_\Tt \log p_{\y_\S|\tt}(\y;\Tt))\\&\qquad\qquad\qquad(\nabla_\Tt \log p_{\y_\S|\tt}(\y;\Tt))^\top]+\I_x,
\end{aligned}
\end{equation}
where 
\[\I_x = \E_{\y_\S,\tt}\left[(\nabla_\Tt \log p_{\tt}(\Tt))(\nabla_\Tt \log p_{\tt}(\Tt))^\top\right]\] 
is the prior Fisher information on $\tt$ (e.g., if $p_\tt(\Tt) = \N(\mathbf{0},\P)$ then $\I_x =\P^{-1}$). Note that the conditional distribution $p_{\y_\S|\tt}(\y;\Tt)$ is normal $\N(\uu_\tt,\G)$. Therefore, 
\begin{equation}
\nabla_\Tt\log p_{\y_\S|\tt}(\y;\Tt) = -(\nabla_\Tt\uu_\S)\G_\S^{-1}(\y_\S-\uu_\S),
\end{equation}
where $[\nabla_\Tt\uu_\S]_i = \X_i\tt+\z_i.$
Using this result we obtain
\begin{equation}\label{eq:bs2}
\begin{aligned}
\B_\S^{-1} & =  \E_{\y_\S,\tt}[(\nabla_\Tt\uu_\S)\G_\S^{-1}(\y_\S-\uu_\S)\\&\qquad\qquad\qquad(\y_\S-\uu_\S)^{\top}\G_\S^{-1}(\nabla_\Tt\uu_\S)^{\top}]+\I_x\\
&\stackrel{(a)} = \E_{\y_\S,\tt}[\E_{\y_\S|\tt}[(\nabla_\Tt\uu_\S)\G_\S^{-1}(\y_\S-\uu_\S)\\&\qquad\qquad\qquad(\y_\S-\uu_\S)^{\top}\G_\S^{-1}(\nabla_\Tt\uu_\S)^{\top}]]+\I_x\\
&=\E_{\y_\S,\tt}[(\nabla_\Tt\uu_\S)\G_\S^{-1}\E_{\y_\S|\tt}\left[(\y_\S-\uu_\S)(\y_\S-\uu_\S)^{\top}\right]\\&\qquad\qquad\qquad\G_\S^{-1}(\nabla_\Tt\uu_\S)^{\top}]+\I_x\\
&\stackrel{(b)} =\E_{\y_\S,\tt}\left[(\nabla_\Tt\uu_\S)\G_\S^{-1}\G_\S\G_\S^{-1}(\nabla_\Tt\uu_\S)^{\top}\right]+\I_x \\
&= \E_{\y_\S,\tt}\left[(\nabla_\Tt\uu_\S)\G_\S^{-1}(\nabla_\Tt\uu_\S)^{\top}\right]+\I_x\\
&=\E_{\y_\S,\tt}\left[\sum_{i\in\S} \frac{1}{\sigma^2_i}(\X_i\tt+\z_i)(\X_i\tt+\z_i)^\top\right]+\I_x\\
&=\sum_{i\in\S} \frac{1}{\sigma^2_i}(\E_{\y_\S,\tt}\left[\X_i\tt\tt^\top\X_i^\top\right]+\E_{\y_\S,\tt}\left[2\X_i\tt\z_i^\top\right]\\&\qquad\qquad\qquad+\E_{\y_\S,\tt}\left[\z_i\z_i^\top\right])+\I_x\\
&=\sum_{i\in\S} \frac{1}{\sigma^2_i}\left(\X_i\P\X_i^\top+\z_i\z_i^\top\right)+\I_x
\end{aligned}
\end{equation}
where to obtain $(a)$ we use the law of total expectation, $(b)$ follows by the definition of covariance matrices (see (2) in the paper), and the last equality follows since we assumed $\E[\tt] = \mathbf{0}$ and $\Cov(\tt) = \E[\tt\tt^\top] = \P$. Inverting the last line that consists of an invertible positive definite matrix establishes the stated results which in turn completes the proof.
\section*{Proof of Theorem 3}
The marginal gain of adding a new observation to a subset $\S$ is
\begin{equation}
\begin{aligned}
f^T_j(\S) &= \mathrm{Tr}\left(\B_{\S\cup\{j\}}^{-1}\right) - \mathrm{Tr}\left(\I_x\right)-\mathrm{Tr}\left(\B_\S^{-1}\right) +\mathrm{Tr}\left(\I_x\right) \\
&= \Tr\left(\B_{\S\cup\{j\}}^{-1}-\B_\S^{-1}\right)\\
&=\Tr\left(\frac{1}{\sigma^2_j}\left(\X_j\P\X_j^\top+\z_j\z_j^\top\right)\right).
\end{aligned}
\end{equation}
Therefore, the marginal gain is trace of a positive semi-definite matrix and hence $f^T_j(\S)\geq 0$ and the function is monotone. Furthermore, since the marginal gain does not depend on set $\S$ it is a modular function.
\section*{Proof of Theorem 4}
Let $\I_j = \frac{1}{\sigma^2_j}\left(\X_j\P\X_j^\top+\z_j\z_j^\top\right)$.
The marginal gain of adding a new observation to a subset $\S$ is
\begin{equation}
\begin{aligned}
f^D_j(\S) &= \log\det\left(\B_{\S\cup\{j\}}^{-1}\right) - \log\det\left(\I_x\right) - 
\log\det\left(\B_\S^{-1}\right) \\&\qquad\qquad\qquad+ \log\det\left(\I_x\right) \\
&= \log\det\left(\B_\S^{-1}
+ \I_j\right)- \log\det\left(\B_\S^{-1}\right) \\
&\stackrel{(a)}{=} \log \frac{\det\B_\S^{-1} 
	\det \left(\mathbf{I} +  \B_\S^{1/2} \I_j \B_\S^{1/2}\right)}
{\det\B_\S^{-1}} \\
&= \log\det \left(\mathbf{I} + \B_\S^{1/2} \I_j \B_\S^{1/2}\right) \\
&\stackrel{(b)}{\ge} 0 ,
\end{aligned}
\end{equation}
where $(a)$ follows from the fact that $\det \left(\A+\B\right) = \det \left(\A\right) \det \left(1+\A^{-1/2}\B\A^{-1/2}\right)$, according to Sylvester's determinant identity, for any positive definite matrix $\A$ and Hermitian matrix $\B$ \cite{bellman1997introduction}, and $(b)$ holds due to $\det\left(\mathbf{I}+\A\right) \ge \left(1+\det \A\right)$ for any positive semi-definite matrix $\A$. Therefore $f^D$ is monotonically increasing. 

Now consider $\S \subseteq \T \subset \Xs$ and $j \in \Xs \backslash \T$. Using the Sylvester's determinant identity we obtain
\begin{equation}
\begin{aligned}
f^D_j(\T)/f^D_j(\S) &=  
\frac{\log\det \left(\mathbf{I} +  \B_\T^{1/2} \I_j \B_\T^{1/2}\right)}{\log\det \left(\mathbf{I} +  \B_\S^{1/2} \I_j \B_\S^{1/2}\right)}
\le 1.
\end{aligned}
\end{equation}
Hence, $c_{f^D}={\max_{(\S,\T,j)\in \tilde{\Xs}}{f^D_j(\T)\slash f^D_j(\S)}} \le 1$ which in turn proves submodularity of D-optimality. 
\section*{Proof of Theorem 5}
The marginal gain of adding a new observation to the subset $\S$ is
\begin{equation}
\begin{aligned}
f^E_j(\S) &= \lambda_{\min}\left(\B_{\S \cup \{j\}}^{-1}\right) -\lambda_{\min}\left(\I_x\right) 
- \lambda_{\min}\left(\B_\S^{-1}\right)\\&\qquad\qquad\qquad + \lambda_{\min}\left(\I_x\right) \\
&= \lambda_{\min}\left(\B_\S^{-1} + \I_j\right) - \lambda_{\min}\left(\B_\S^{-1}\right) \\
&\stackrel{(a)}{\ge} \lambda_{\min}\left(\I_j\right)
\end{aligned}
\end{equation}
where $(a)$ follows from $\lambda_{\min}(\A+\B) \ge \lambda_{\min}(\A) + \lambda_{\min}(\B)$ according to Weyl's inequality for two Hermitian matrices \cite{bellman1997introduction}. The positive semi-definiteness of $\I_j$ implies $f^E_j(\S) \ge 0$ and hence, monotonicity of $f^E$ is established.

We now provide bounds on additive and multiplicative weak-submodularity constants of $f^E(\S)$
(Note that it can be shown using simple examples that $f^E$ is not in general weak submodular). Let $\S \subseteq \T \subset \Xs$ and $j \in \Xs \backslash \T$.
\begin{equation}
\begin{aligned}
f^E_j(\T)/f^E_j(\S) &= 
\frac{\lambda_{\min}\left(\B_\T^{-1} + \I_j\right) - \lambda_{\min}\left(\B_\T^{-1}\right)}
{\lambda_{\min}\left(\B_\S^{-1} + \I_j\right) - \lambda_{\min}\left(\B_\S^{-1}\right)} \\
&\stackrel{(b)}{\le} \frac{\lambda_{\min}\left(\B_\T^{-1}\right) + \lambda_{\max}\left(\I_j\right) - \lambda_{\min}\left(\B_\T^{-1}\right)}
{\lambda_{\min}\left(\B_\S^{-1}\right) + \lambda_{\min}\left(\I_j\right) - \lambda_{\min}\left(\B_\S^{-1}\right)} \\
&\le \frac{\lambda_{\max}\left(\I_j\right)}{\lambda_{\min}\left(\I_j\right)} ,
\end{aligned}
\end{equation}
where $(b)$ follows from Weyl's inequality \cite{bellman1997introduction}. Therefore,
\begin{equation}
c_{f^E}={\max_{(\S,\T,j)\in \tilde{\Xs}}{f^D_j(\T)\slash f^D_j(\S)}} \le 
\max_{j \in \Xs} \frac{\lambda_{\max}\left(\I_j\right)}{\lambda_{\min}\left(\I_j\right)}.
\end{equation}
For the additive weak-submodularity constant, we have
\begin{equation}
\begin{aligned}
f^E_j(\T) - f^E_j(\S) &= \lambda_{\min}\left(\B_\T^{-1} + \I_j\right) - \lambda_{\min}\left(\B_\T^{-1}\right)\\&\qquad- \lambda_{\min}\left(\B_\S^{-1} + \I_j\right) + \lambda_{\min}\left(\B_\S^{-1}\right) \\
&\stackrel{(c)}{\le} \lambda_{\max}\left(\I_j\right) - \lambda_{\min}\left(\I_j\right)
\end{aligned}
\end{equation}
where $(c)$ follows from Weyl's inequality. Hence,
\begin{equation}
\begin{aligned}
\epsilon_{f^E}&={\max_{(\S,\T,j)\in \tilde{\Xs}}{f^D_j(\T) - f^D_j(\S)}} \\&\le 
\max_{j \in \Xs} \left(\lambda_{\max}\left(\I_j\right) - \lambda_{\min}\left(\I_j\right)\right).
\end{aligned}
\end{equation}
\section*{Proof of Theorem 6}
We first prove the monotonicity. Let $\I_j = \frac{1}{\sigma^2_j}\left(\X_j\P\X_j^\top+\z_j\z_j^\top\right)$. For any set $\S$ and $j\in \Xs\backslash\S$, define 
\begin{equation}
\tilde{\F}_{\S,j} = \I_x+\sum_{i\in\S} \I_i+\sigma^2_j\I_j = {\F}_{\S}+\sigma^2_j\I_j,
\end{equation}
where both $\tilde{\F}_{\S,j}$ and $\F_{\S}$ are invertible and positive definite (PSD) matrices. Using the matrix inversion lemma \cite{bellman1997introduction} as well as some algebraic simplifications, we obtain an expression for the marginal gain according to 
\begin{equation}
\begin{aligned}
f_j^A(\S)& = \frac{\z_j^\top\tilde{\F}_{\S,j}^{-2}\z_j}{\sigma_j^2+\z_j^\top\tilde{\F}_{\S,j}^{-1}\z_j}\\&+\Tr\left(\F_{\S}^{-1}\X_j(\sigma_j^2\P^{-1}+\X_j^\top\F_{\S}^{-1}\X_j)^{-1}\X_j^\top\F_{\S}^{-1}\right).
\end{aligned}
\end{equation}
Notice the first term on the right-hand side is positive since $\tilde{\F}_{\S,j}$ is PSD and hence the quadratic form $\z_j^\top\tilde{\F}_{\S,j}^{-2}\z_j$ is also positive. Further, The second term on the right-hand side is also positive as it is trace of the quadratic form $\F_{\S}^{-1}\X_j(\sigma_j^2\P^{-1}+\X_j^\top\F_{\S}^{-1}\X_j)^{-1}\X_j^\top\F_{\S}^{-1}$ which is also PSD because the matrix $(\sigma_j^2\P^{-1}+\X_j^\top\F_{\S}\X_j)^{-1}$ is itself PSD. Thus, the marginal gain is positive and the function is monotonically increasing.

We now provide bounds on additive and multiplicative weak-submodularity constants of $f^E(\S)$
(Note that it can be shown $f^E$ is not in general submodular). Finding these bounds in the general form form of model requires intense algebraic techniques and the resulting bounds will not be interpretable. In stead, we here provide bounds in scenarios where $\z_i=\mathbf{0}$ and $\X_i = \x_i\x_i^\top$ (rank 1) which is motivated by the phase retrieval applications. In this setting, it can be shown the marginal gain simplifies to 
\begin{equation}
f_j^A(\S) = \frac{\x_j^\top\tilde{\F}_{\S}^{-2}\x_j}{\x_j^\top(\sigma_j^2\P+\tilde{\F}_{\S}^{-1})\x_j},
\end{equation}
where $\tilde{\F}_{\S} = \I_x+\sum_{i\in\S}\frac{1}{\sigma_j^2}\x_i\x_i^\top\P\x_i\x_i^\top$. Hence, the definition of multiplicative weak-submodularity constant yields,
\begin{equation}
\begin{aligned}
c_{f^A}&={\max_{(\S,\T,j)\in \tilde{\Xs}}{f^A_j(\T)\slash f^A_j(\S)}}\\
&= \max_{(\S,\T,j)\in \tilde{\Xs}}\frac{(\x_j^\top\tilde{\F}_{\T}^{-2}\x_j)(\x_j^\top(\sigma_j^2\P+\tilde{\F}_{\S}^{-1})\x_j)}{(\x_j^\top\tilde{\F}_{\S}^{-2}\x_j)(\x_j^\top(\sigma_j^2\P+\tilde{\F}_{\T}^{-1})\x_j)}\\
&\leq \max_{(\S,\T,j)\in \tilde{\Xs}} \frac{\lambda_{\max}(\tilde{\F}_{\T}^{-2})\lambda_{\max}(\sigma_j^2\P+\tilde{\F}_{\S}^{-1})}{\lambda_{\min}(\tilde{\F}_{\S}^{-2})\lambda_{\min}(\sigma_j^2\P+\tilde{\F}_{\T}^{-1})},
\end{aligned}
\end{equation}
where the last inequality follows from the Courant–Fischer min-max theorem \cite{bellman1997introduction}. Notice that by Weyl's inequality $\lambda_{\max}(\tilde{\F}_\S^{-1})=\lambda_{\min}(\tilde{\F}_\S)^{-1}$ and $\lambda_{\min}(\tilde{\F}_\T)\geq\lambda_{\min}(\tilde{\F}_\S)\geq\lambda_{\min}(\tilde{\F}_\emptyset)=\lambda_{\min}(\I_x)$. Therefore,
\begin{equation}
\begin{aligned}
c_{f^A}&\leq\max_j\frac{\lambda_{\max}(\I_x^{-1})^{2}\lambda_{\max}(\sigma_j^2\P+\I_x^{-1})}{\lambda_{\min}(\tilde{\F}_{[n]}^{-1})^{2}\lambda_{\min}(\sigma_j^2\P+\tilde{\F}_{[n]}^{-1})}\\
&\leq \max_j\frac{\lambda_{\max}(\I_x^{-1})^{3}(\frac{\lambda_{\max}(\sigma_j^2\P)}{\lambda_{\max}(\I_x^{-1})}+1)}{\lambda_{\min}(\tilde{\F}_{[n]}^{-1})^{3}(\frac{\lambda_{\min}(\sigma_j^2\P)}{\lambda_{\min}(\tilde{\F}_{[n]}^{-1})}+1)}.
\end{aligned}
\end{equation}
Noting $\tilde{\F}_{[n]}^{-1} = \B_{[n]}$ completes the proof of bounded $c_{f^A}$. We can also use more applications of Weyl's inequality to achieve looser yet more intuitive and compact bounds. Using similar techniques such as applications of Courant–Fischer min-max theorem and Weyl's inequality we obtain the stated results for $\epsilon_{f^A}$.

\end{document}